\documentclass[showpacs,pre,aps,twocolumn]{revtex4}
\usepackage{graphicx,epsfig}
\begin{document}
\title{Transition to Turbulence in \\ Coupled Maps on Hierarchical Lattices} 
\author{M. G. Cosenza}
 \email{mcosenza@ciens.ula.ve}
\author{K. Tucci}
 \email{kay@ula.ve}

 \affiliation{
        $^*$Centro de Astrof\'{\i}sica Te\'orica, Facultad de
             Ciencias, Universidad de Los Andes,
        Apartado Postal 26 La Hechicera, M\'erida 5251, Venezuela \\
        $^\dag$SUMA - CeSiMo, Universidad de Los Andes, M\'erida, Venezuela}

\begin{abstract}
General hierarchical lattices of coupled maps are considered as 
dynamical systems. These models may describe many processes occurring in
heterogeneous media with tree-like structures. The transition to turbulence
via spatiotemporal intermittency is investigated for these geometries.  
Critical exponents associated to the
onset of turbulence are calculated as functions of the parameters of the
systems. No evidence of nontrivial collective behavior is observed in the
global quantity used to characterize the spatiotemporal dynamics.
\end{abstract}

\maketitle

\section{Introduction.}
Coupled map lattices are spatiotemporal
dynamical systems comprised of an interacting array
of discrete-time maps.  Coupled map lattices have provided fruitful models for the study of
many spatiotemporal processes in a variety of contexts \cite{Survey}.
In most of these studies, the dynamical processes have been assumed to
take place on uniform Euclidean spaces.   
However; because of their discrete spatial nature, coupled map lattice systems seem especially 
appropriate for investigating physical phenomena occurring in heterogeneous
media. In this respect, there are some recent models of coupled maps defined on fractal
lattices \cite{Us1,Us2}, and also
an investigation of coupled maps on a Cayley tree \cite{Gade}.

In this article we consider coupled maps defined on generalized hierarchical 
lattices, or multi-branching trees, as dynamical systems. Examples of
phenomena where hierarchical structures appear include DLA
clusters, capillarity, chemical reactions in porous media \cite{Kopel},
turbulence \cite{Sree}, ecological systems \cite{Hogg},
interstellar cloud complexes \cite{Scalo}, etc.   
Hierarchical structures have also been studied in 
neural nets, because of their exponentially large storage capacity
\cite{Admit}.
Although many hierarchical structures found in nature have random 
ramifications, here we study the case of simple, deterministic
tree-like arrays. This allows to focus on the changes induced in
spatiotemporal processes as a result of the hierarchical structure
of the interactions in the system.

In Sec.~II, coupled map lattice models for the treatment of generalized hierarchical lattices
are presented.   
As an application, in Sec.~III we study the phenomenon of spatiotemporal 
intermittency
as a mechanism for the transition to turbulence in coupled maps on
hierarchical lattices.
The model is based on the one introduced earlier by Chat\'{e} and
Manneville for regular Euclidean lattices in one and two dimensions \cite{Chate1},
and which captures the essential features of the transition to turbulence
in extended systems. A discussion is given in Sec.~IV. 

\section{Coupled maps on hierarchical lattices.}
Generalized trees constitute a class of 
hierarchical lattices which can be generated for any constant ramification 
number $k$. At the initial level (which we call level $0$), there is one cell
which splits into $k$ branches connecting $k$ daughters cells, which
constitute the
level of construction $1$. Each cell then splits into $k$ daughters, producing
$k^2$ sites at level $2$. This construction continues until some level $n$.
There are $k^m$ cells at the level of construction $m$; $m=0,1\ldots,n$.
Thus, each cell in the lattice has one parent and $k$ daughters, except for
the level $0$ cell, which has no parent, and for boundary cells at level $n$,
which have no daughters. The number of cells lying on the boundary is $k^n$.

Each cell in a tree constructed until level $n$ can be identified by 
an index $i=1,2,\ldots,N$, where $N$ is the total number of cells on 
the tree,
or the system size, given by $N=(k^{n+1}-1)/(k-1)$.

\begin{figure}[h]
\epsfig{file=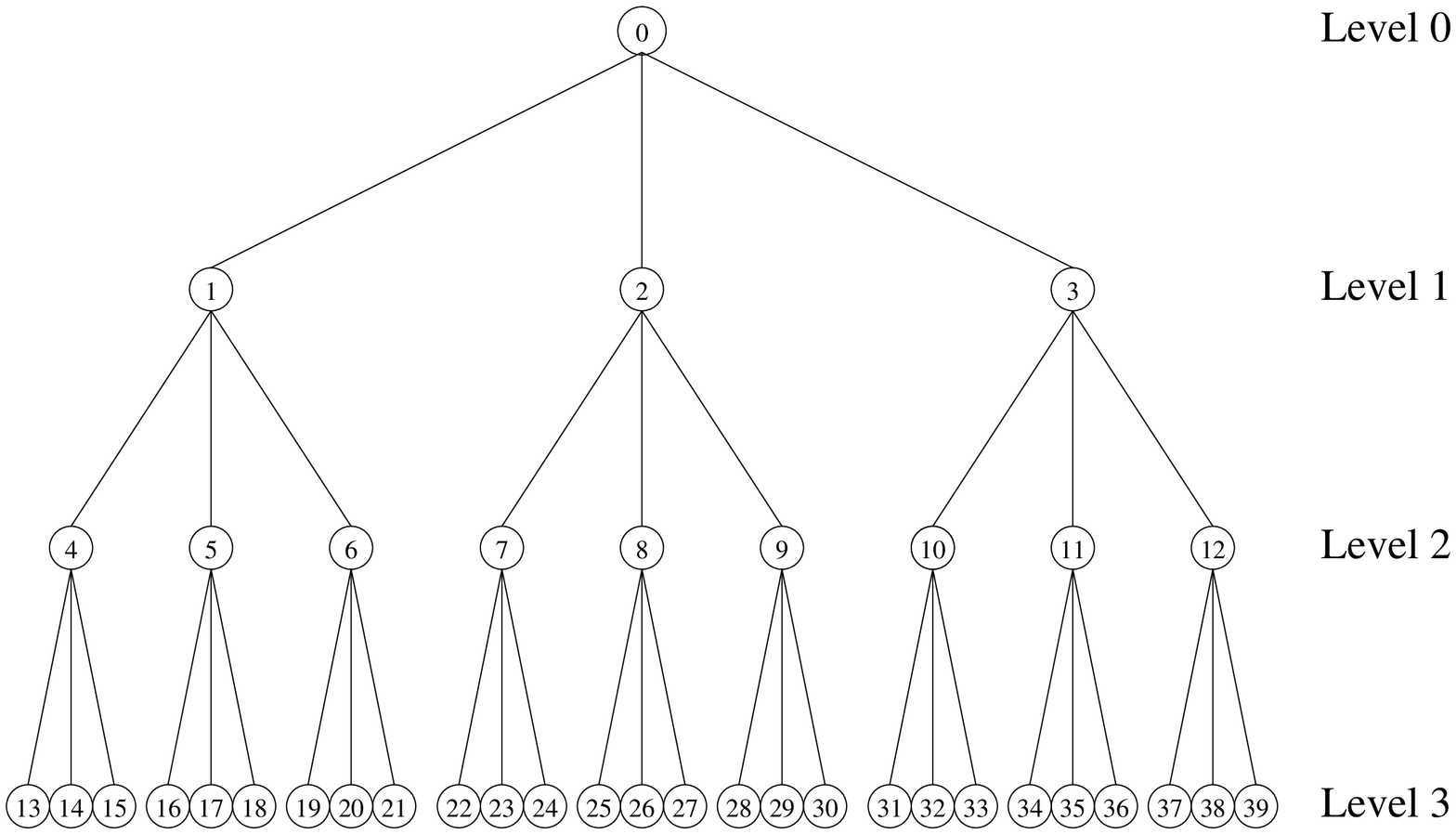,width=0.2\textwidth,angle=270,clip=}
\caption{ Hierarchical lattice with ramification $k=3$ and level of
construction $n=3$, showing the indices on the cells.
\label{fig1}}
\end{figure}

A cell denoted by the index $i$ at a level $m$ is coupled only to neighbor cells 
belonging to
adjacent levels 
on the lattice, {\it i.e.}, to its parent and to its daughters, which we
indicate by indexes $i_p$ and $i_d$, respectively.
We do not 
consider interactions between cells belonging to the same level. 
Thus, a discrete diffusive coupling can be defined
according to the following scheme. If $i=0$, then the cell is at the level $0$
of the network and it is coupled to its $k$ daughters, which have the indices
$i=1,2\ldots,k$. 
Cells with $i>0$ are coupled to cells with indices $i_p$ and $i_d$, such as 
\begin{equation}
i_p= \mbox{int}\left(\frac{i-1}{k}\right),
\end{equation}
where the function {\sl int} means integral part,
and
\begin{equation}
i_d=(i \times k)+j; \quad  j=1,2,\ldots,k; \quad \mbox{if} \;\; i<(N-k^n), 
\end{equation}
and 
$i_d$ is not defined for boundary cells, which have $i\ge (N-k^n)$.
The level $0$ cell has $k$ neighbors, intermediate cells have $k+1$ neighbors,
and boundary cells for which $i>(N-k^n)$ possess only one neighbor.
As an illustration, Fig.~(\ref{fig1}) shows a hierarchical tree with ramification 
number $k=3$
and construction level $n=3$. The lattice size is $N=40$. The indices on each
cell are indicated. In this example, the cell labeled by $i=11$ belongs to 
level $n=2$; its parent has index $i_p=3$, and its daughters are labeled by 
$i_d=34,35$, and $36$, respectively.

The state of each site on the lattice can be assigned a continuous variable $x$, 
which evolves 
according to a deterministic rule depending on its own value and the values of
its connecting neighbors.
The equations describing the dynamics of a diffusively coupled map
hierarchical lattice at level of construction $n$ can be written as

\noindent{(a)} level $0$ cell:
\begin{equation}
\label{eq0}
x_{t+1}(0)= (1-k \gamma) f(x_t(0))+\gamma 
\sum_{j\in i_d}f\left(x_t(j)\right);
\end{equation}
(b) for $i<(N-k^n)$,
\begin{equation}
\label{ec1}
\begin{array}{ll}
x_{t+1}(i)= & \left[1-(k+1)\gamma \right]f(x_t(i))+ \cr
            & \gamma f\left(x_t(i_p)\right)+ 
  \gamma \sum_{j\in i_d} f\left(x_t(j)\right);
\end{array}
\end{equation}
(c) and for boundary cells with $i\ge (N-k^n)$,
\begin{equation}
x_{t+1}(i)= f\left(x_t(i)\right)+ \gamma f\left(x_t(i_p)\right);
\end{equation}
where $x_t(i)$ gives the state of the cell 
$i$ at discrete time $t$; 
$i_p$ and $i_d$ label the $N$ cells on the lattice;
$\gamma$ is a parameter measuring the coupling strength between neighboring
sites, and  
$f(x)$ is a nonlinear function describing the local dynamics. The above 
coupled map lattice
equations can be generalized to include nonuniform coupling and/or varying
ramifications within the network. By providing appropriate local dynamics and
couplings, different spatiotemporal phenomena can be studied on tree-like
structures.

\section{An application: transition to turbulence.}
Spatiotemporal intermittency in extended systems consists of
a sustained regime where coherent and chaotic domains coexist and evolve in
space and time.
The transition to turbulence via spatiotemporal intermittency has been studied
in coupled map lattices whose spatial supports are Euclidean
\cite{Chate1,Kan,Stassi} and also in fractal lattices \cite{Us2}. 
A local map possessing the minimal requirements for observing 
spatiotemporal intermittency is \cite{Chate1}
\begin{equation}
\label{map}
f(x)=\left\{
\begin{array}{ll}
\frac{r}{2}\left(1-\left| 1-2x \right|\right), & \mbox{if $x \in [0,1]$} \\
x, & \mbox{if $x > 1$},
\end{array}
\right.
\end{equation}
with $r>2$. This map is chaotic for
$f(x)$  in $[0,1]$. However, for $f(x) >1$ the iteration is
locked on a fixed point. The local state can thus be seen as a continuum
of stable ``laminar" fixed points $(x>1)$ adjacent to a chaotic repeller
or ``turbulent" state $(x \in [0,1])$. 

\begin{figure}[h]
\epsfig{file=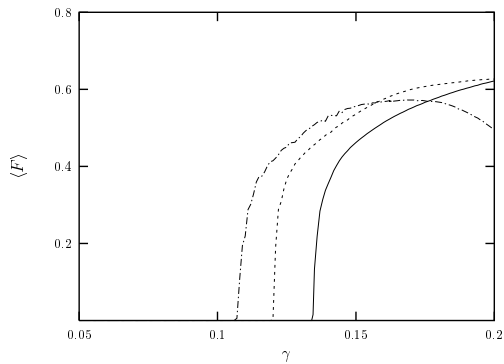,width=0.3\textwidth,angle=90,clip=}
\caption{Mean turbulent fraction $\langle F \rangle$
as a function of the coupling $\gamma$ for different
hierarchical lattices. Thick left curve, $k=2$, $n=12$; 
middle curve, $k=3$,
$n=8$;
thin right curve, $k=4$, $n=7$. The local parameter is fixed at $r=3$.
\label{fig2}}
\end{figure}

In regular dimensions, the turbulent state can propagate through the
lattice in time for a large enough coupling, producing sustained regimes
of spatiotemporal intermittency \cite{Chate1}.
Here, we investigate the phenomenon of transition to turbulence in
hierarchical lattices using
the local map $f$ (Eq.(\ref{map})) in the coupled system
described by Eqs.~(3)-(5). The local parameter is fixed at the value $r=3$ in all
the calculations. As observed for regular lattices, starting from
random initial conditions and after some transient regime, our systems
settle in a stationary statistical behavior. The transition to turbulence can
be characterized through the average value of the instantaneous
fraction of turbulent
sites $F_{t}$, a quantity that serves as the order parameter
\cite{Chate1}.
We have calculated  $\langle F \rangle$ as a function
of the coupling parameter $\gamma$ for several hierarchical lattices from
a time average of the instantaneous turbulent fraction $F_t$, as
\begin{equation}
\label{av}
\langle F \rangle={1 \over T} \sum_{t=1}^T F_t.
\end{equation}
About $10^4$ iterations were discarded before taking the
time average in Eq.~(\ref{av}) and $T$ was typically taken at the value
$10^4$. 
Fig.~(\ref{fig2}) shows $\langle F \rangle$ vs. $\gamma$ for hierarchical
lattices with ramification numbers 
$k=2, 3$ and $4$, and levels of construction $n=12, 8$ and $7$, respectively.
We have verified that increasing the averaging time $T$ or the lattice size
$N$ does not have appreciable effects on our results.
The standard deviations about each value of $\langle F \rangle$ are very
small on the scale of Fig.~(\ref{fig2}).
In fact, the fluctuations of $F_t$ 
become smaller for increasing ramification $k$ and/or for higher levels
of construction. In comparison, high dimensional Euclidean lattices \cite{Chate3}, 
as well as fractal 
lattices \cite{Us2} and globally coupled maps \cite{CP},
exhibit large oscillations in the time series
of the instantaneous turbulent fraction $F_t$
for some parameter values. In those cases, the fluctuations about the mean values of
the turbulent fraction do not decrease with increasing
lattice size or averaging time, pointing to the presence of nontrivial collective behavior in
the systems. In contrast, normal statistical properties are
observed in hierarchical arrays of coupled maps, even in the cases when these lattices
have the same numbers of local connections than the other arrays.
Note that fractal lattices
are also spatially nonuniform. However, the behavior of a statistical
variable such as the turbulent fraction, is different for
hierarchical lattices. Calculations of the mean field
in Cayley trees 
using local logistic maps  
neither show evidence of nontrivial
collective behavior \cite{Gade}. Thus, the topology
of the underlying geometry seems to have a relevant influence on the kind of 
behavior that may
emerge in global quantities in spatiotemporal systems. 

There exists a critical
value of the coupling $\gamma_{c}$ for the onset of
turbulence which becomes smaller for increasing
ramification of the trees.
The threshold for intermittency
is achieved when the probability of reinjection into the unit interval
is balanced by the probability of escape from this interval, which is
constant for fixed $r$. The former is proportional to $\gamma$ and to the
number of coupled neighbors, one of them at least being turbulent.
Thus, $\gamma_{c}$ must decrease with increasing $k$, for fixed $r$.
Fig.~({\ref{fig3}) shows $\gamma_{c}$ as a function of the ramification number $k$,
verifying this assumption. 
\begin{figure}[h]
\epsfig{file=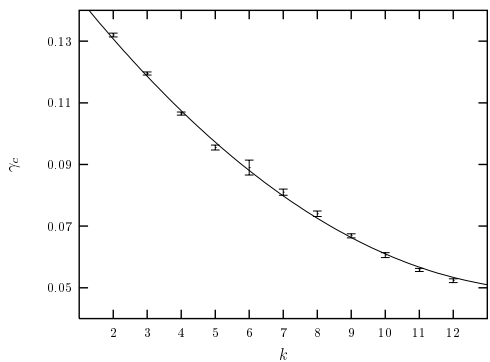,width=0.3\textwidth,angle=90,clip=}
\caption{Critical coupling $\gamma_c$ as a function 
of the ramification $k$.
\label{fig3}}
\end{figure}

As for other lattices, the transition from a laminar regime to turbulence can be characterized
by critical exponents like $\beta$ which scales the variation of the
order parameter near the transition point:
$\langle F \rangle \sim(\gamma-\gamma_{c})^{\beta}$.
We numerically checked this relation. Our results are given in Fig.~(\ref{fig4})
for lattices with different ramification numbers, 
using a log-log plot. 
\begin{figure}[h]
\epsfig{file=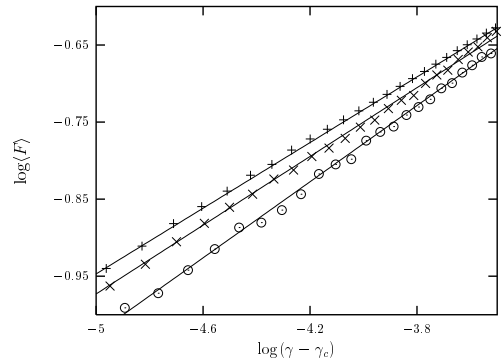,width=0.3\textwidth,angle=90,clip=}
\caption{Log-log plot of $\langle F \rangle$ vs $(\gamma-\gamma_c)$ for
hierarchical lattices: $k=2$ ($+$); $k=3$ ($\times$); $k=4$ ($\circ$).
\label{fig4}}
\end{figure}
Table~(\ref{tab1}) gives the critical exponents $\beta$ calculated from the slopes of
Fig.~(\ref{fig4})
for different ramifications $k$, with their respective errors.
The values of $\beta$ for these hierarchical lattices with different
ramifications are of the order of that
of a one-dimensional lattice, which is $\beta=0.250 \pm 0.005$ 
\cite{Chate1}.
Thus, the
statistical properties of the transition to turbulence for
hierarchical lattices are closer to that of a one-dimensional lattice in
this scenario. It should also be noticed that 
the global dynamics of
one-dimensional coupled map
lattices neither display nontrivial collective behavior \cite{Chate4}. 
\begin{table}[h]
  \begin{tabular}{|c|c|c|}
  \hline
    {\bf $k$ } & {\bf $n$ } & {\bf $\beta$ } \\
  \hline
    $\ \quad 2\ \quad $& $\ \quad 13\ \quad $& $\ \quad 0.21 \pm 0.01\ \quad  $ \\
    $3$& $8$& $0.22 \pm 0.02 $ \\
    $4$& $7$& $0.25 \pm 0.03 $ \\
    $5$& $7$& $0.28 \pm 0.02 $ \\
  \hline
  \end{tabular}
\caption{Critical exponents $\beta$  
for hierarchical lattices with different ramifications $k$. The 
corresponding levels of
construction $n$ are also indicated.
\label{tab1}}
\end{table}

\section{Discussion.}
We have introduced coupled map lattice systems with hierarchical couplings.
These models are especially suited for investigating spatiotemporal phenomena
occurring in heterogeneous media with an underlying tree-like geometry,
and they may also represent the hierarchical interactions taking place in
some organized systems.

Although we employed 
the simple Chat\'e-Manneville map for illustrating 
the transition to turbulence 
on hierarchical lattices, more sophisticated coupled map models
may prove useful in the study of fully developed turbulence in fluids, where
dissipation of energy occurs in hierarchical cascades from larger to smaller
structures \cite {Sree}. 

In hierarchical lattices, as defined in Sec.~II, there exists a unique
path connecting any two given elements on the network. This same topological
property occurs
in a one-dimensional lattice with open ends. 
The similarity in the scaling behavior at the transition to turbulence
between hierarchical lattices of different ramifications
and a one-dimensional lattice suggests that topological properties
of the underlying lattice are relevant in the
global behavior of spatiotemporal systems.

In addition, 
the global quantity Eq.~(\ref{av})
follows a normal statistical behavior in hierarchical lattices, while nontrivial collective 
oscillations arise in other geometries.   
Hierarchical lattices have different topological properties than those
of regular higher dimensional Euclidean arrays or fractal lattices
which have previously been
considered as coupled map systems. No closed loops exist in hierarchical
lattices, and the number of local connections is not uniform since an
appreciable fraction of cells possessing only one neighbor lie on
the boundary.  
As it has been suggested 
\cite{Us2,Chate4,Mc}, the topology of the lattice seems to play a more important role
on the emergence of collective behaviors than the number of local connections or the
dimensionality of the space. The failure to observe collective behavior in
our models also points to this conclusion.       
Future work on hierarchical and other nonuniform lattices may contribute to
answer this and other remaining questions on the collective dynamics of chaotic
extended systems. 

\section*{Acknowledgment}
This work was supported by the Consejo de Desarrollo
Cient\'{\i}fico, Human\'{\i}stico y Tecnol\'ogico of
the Universidad de
Los Andes, M\'erida, Venezuela.


\begin{references}
\bibitem{Survey} {\sl Theory and Applications of
Coupled Map Lattices}, {\sl Chaos} {\bf 2}, No. 3, 
edited by K. Kaneko, Wiley, N. Y., (1993).
\bibitem{Us1} M. G. Cosenza and R. Kapral, {\sl Phys. Rev. A} 
{\bf 46}, 1850 (1992);
{\sl Chaos} {\bf 2}, 329 (1992). 
\bibitem{Us2} M. G. Cosenza and R. Kapral, {\sl Chaos} {\bf 4}, 99 (1994).
\bibitem{Gade} P. M. Gade, H. Cerdeira, and R. Ramaswamy, {\sl Phys. Rev. E}
{\bf 52}, 2478 (1995).
\bibitem{Kopel}
{\sl The Fractal Approach to Heterogeneous
Chemistry}, edited by D. Avnir, Wiley, New York, 1989.
\bibitem{Sree} C. Meneveau and K. R. Sreenivasan, {\sl Phys. Rev. Lett.} {\bf 59},
1424 (1987).
\bibitem{Hogg} T. Hogg, B. A Huberman and J. McGlade, {\sl Proc. R. Soc. London}
{\bf B 237}, 43 (1989).
\bibitem{Scalo} P. Houlahan and J. Scalo, {\sl Ap. J.} {\bf 393}, 172 (1992).
\bibitem{Admit} D. Admit, {\sl Modeling Brain Function: The World of
Attractor Neural Nets}, Cambridge U. Press, Cambridge, 1989.
\bibitem{Chate1} H. Chat\'{e} and P. Manneville, {\sl Physica D} {\bf 32},
409 (1988); {\sl Europhys. Lett.} {\bf 6}, 591 (1988).
\bibitem{Kan} K. Kaneko, {\sl Prog. Theor. Phys.} {\bf 74}, 1033 (1985).
\bibitem{Stassi} D. Stassinopoulos and P. Alstr{\o}m, {\sl Phys. Rev. A}
{\bf 45}, 675 (1992).
\bibitem{Chate3} H. Chat\'e and P. Manneville, in {\sl New Trends in
Nonlinear Dynamics and Pattern Forming Phenomena}, edited by
P. Coullet and P. Heurre, Plenum, New York (1989).
\bibitem{CP} M. G. Cosenza and A. Parravano, {Phys. Rev. E} {\bf 53},
6032 (1998).
\bibitem{Chate4} H. Chat\'{e} and P. Manneville, {\sl Prog. Theor. Phys.}
{\bf 87}, 1 (1992); {\sl Chaos} {\bf 2}, 307 (1992).
\bibitem{Mc} M. G. Cosenza, {\sl Physica A} {\bf 257},
357 (1998).
\end{references}
\end{document}